\documentclass[12pt]{article}
\usepackage[dvips]{epsfig}

\newcommand{\be}{\begin{equation}}
\newcommand{\ee}{\end{equation}}
\newcommand{\bn}{\begin{eqnarray}}
\newcommand{\en}{\end{eqnarray}}

\newcommand{\nn}{\nonumber}

\def\bea{\begin{eqnarray}}
\def\eea{\end{eqnarray}}
\def \ps
{p\!\!\!/}
\def \pa
{A\!\!\!/}
\def \pb
{B\!\!\!/}

\newcommand{\beq}{\begin{equation}}
\newcommand{\eeq}{\end{equation}}
\newcommand{\no}{\noindent }
\begin{document}

\title{\textbf{Perturbative bosonization from two-point correlation functions}}
\author{D. Dalmazi, A. de Souza Dutra and Marcelo Hott \\
\textit{{UNESP - Campus de Guaratinguet\'a - DFQ} }\\
\textit{{Av. Dr. Ariberto Pereira da Cunha, 333} }\\
\textit{{CEP 12516-410 - Guaratinguet\'a - SP - Brazil.} }\\
\textsf{E-mail: dalmazi@feg.unesp.br, dutra@feg.unesp.br }\\
\textsf{and hott@feg.unesp.br}}
\date{\today}
\maketitle

\begin{abstract}
Here we address the problem of  bosonizing massive fermions without making expansions
in the fermion masses in both massive $QED_2$ and $QED_3$ with $\, N \, $ fermion
flavors including also a  Thirring coupling. We start from two point correlators
involving the $U(1)$ fermionic current and the gauge field. From the tensor structure
of those correlators we prove that the $U(1)$  current must be identically conserved
(topological)  in the corresponding bosonized theory both in $D=2$ and $D=3$
dimensions. We find an effective generating functional in terms of bosonic fields
which reproduces those two point correlators and from that we obtain a map of the
Lagrangian density $\bar{\psi}^{r}\,(i\,\partial \!\!\!/\,-\,m\,){\psi}^{r}$  into a
bosonic one in both dimensions. This map is nonlocal but it is independent of the
eletromagnetic and Thirring couplings, at least in the quadratic approximation for the
fermionic determinant.

\textit{{PACS-No.:} 11.15.Bt , 11.15.-q }
\end{abstract}



\newpage

\section{Introduction}


One among many dreams of the theoretical physicists nowadays is the possibility of
extending to higher dimensions ($D>2$) the bosonization of fermionic models. This can
be justified by some good properties of one formulation  in contrast with the other
one. For instance, strong coupling physics in one model corresponds to weak coupling
in the other one. A classic example is the map between the massive Thirring and
Sine-Gordon models in $D=2$.  Another interesting aspect of such map is the fact that
the usual electromagnetic charge in the fermionic model corresponds to the topological
charge of the associated soliton field. Furthermore, we can have  a map between a
linear theory like massive free fermions and a nonlinear one
 (Sine-Gordon at $\beta ^{2}=4\,\pi $) on the other side.

In view of these and other interesting properties a lot of work has been devoted to
the issue of bosonization \cite {Jackiw,Coleman,mandelstam} ( see also
\cite{livroElcio} ). There has also been many attempts to generalize those ideas to
higher dimensions [5-22]. For massive fermions in $\, D=2 \, $ most of the methods is
based on expansions around massless fields which are local conformal theories. In
$D=3$, although the case of massless free fermions can still be mapped into a bosonic
theory ( see \cite{marino} ) such theory is nonlocal. Besides, the conformal group is
finite in $\, D=3\, $ and not so powerful as in $\, D=2 \, $ which makes expansions
around the massless case nontrivial. The other possibility is to employ functional
methods. Once again the case of massless fermions is easier to deal with since the
fermionic determinant can be exactly calculated for $m=0$. For massive fermions in
$D=2$ a nontrivial Jacobian under chiral transformations plays a key role in deriving
the Sine-Gordon model ( see [26-30]). In $\, D=3\, $, chiral transformations play no
role and although some nonperturbative information is known \cite{linhares} about the
fermionic determinant we are basically left with approximate methods like the one used
in this work. On one hand, the method used here is inspired in the approach carried
out in \cite{bm2}
 which is rather simple and based on two point correlation functions. On the other hand,
 our bosonization rules departure from those in \cite{bm2} in the sense
that they are independent of the interactions.  We  extend the approach of \cite{bm2}
by introducing an eletromagnetic interaction and making  use of the $1/N$ expansion
which allows us to go beyond the lowest-order in the coupling constants. In the next
section we introduce the model we are working with and obtain a general expression for
the generating funtional of the current and gauge field correlators. The expression is
valid for arbitrary dimensions and depends on the vacuum polarization tensor. In
sections 3 and 4 we make the calculations explicit in $D=2$ and $D=3$ dimensions
respectively. We first obtain in the fermionic theory the two point current
correlation functions involving also the gauge field and then we write the current in
terms of bosonic fields and derive the corresponding action for such fields which
reproduces their correlators. In the final section we draw some conclusions and
comment on similar approaches in the literature.

\section{Generating Functionals }

We start by introducing the notation which will be used in both $D=2$ and $%
D=3$. The generating functional for a generalized QED
with Thirring self-interaction is given by
\begin{eqnarray}
Z\left[ J^r_{\mu },K_{\mu }\right] &=&\int \mathcal{D}A_{\mu }\,\prod_{r=1}^{N}\mathcal{D}%
\psi_r \,\mathcal{D}\bar{\psi}_r\,exp\left\{ i\,\int \,d^{D}x\,\,\left[ -\frac{1%
}{4}\,F_{\mu \nu }^{2}\,+\,\bar{\psi}_{r}\,(i\,\partial \!\!\!/\,-\,m\,-\,%
\frac{e}{\sqrt{N}}\,\pa)\psi_{r}\right. \right.  \nonumber \\
&&\left. \left. -\,\frac{g^{2}}{2N}\,(\,\overline{\psi }_{r}\gamma ^{\mu }\psi
_{r})^{2}\, +\frac{\lambda }{2}\left( \partial _{\mu }A^{\mu }\right) ^{2}+\,J_{\mu
}^{r}\left( \overline{\psi }_{r}\gamma ^{\mu }\psi_{r}\right) +\,K_{\mu }\,A^{\mu
}\right] \right\} \ ,
\end{eqnarray}

\noindent where $N$ is the number of fermion flavors and summation over the
repeated flavor index $r$ ($r=1,2,...,N$) is assumed. It is convenient to
introduce an auxiliary vector field $B_{\mu }$ and work with the physically
equivalent generating functional:

\begin{eqnarray}
Z\left[ J^r_{\mu },K_{\mu }\right]  &=&\int \mathcal{D}A_{\mu }\mathcal{D}B_{\mu }
\prod_{r=1}^N\mathcal{D}\psi_r \,\mathcal{%
D}\bar{\psi}_r\,exp\left\{ i\,\int d^{D}x\,\left[ -\frac{1}{4}\,F_{\mu \nu
}^{2}+{\frac{1}{2}}B_{\mu }B^{\mu }+ \frac{\lambda }{2}\left( \partial _{\mu }A^{\mu
}\right)
^{2}+ \right. \,\right.  \nonumber \\
&&\left. \left. +\bar{\psi}_r\,(i\,\partial \!\!\!/\,-\,m\,-\,\frac{e}{%
\sqrt{N}}\,\pa-\,\frac{g}{\sqrt{N}}\,\pb+J\!\!\!/^{r})\psi_r+\,K_{\mu }\,A^{\mu
}\right] \right\} \ ,
\end{eqnarray}

\no Integrating over the fermionic fields we obtain

\begin{eqnarray}
Z \left[ J^r_{\mu },K_{\mu }\right] &=&\int \mathcal{D}A_{\mu }\mathcal{D}B_{\mu
}\exp\left\{ i\,\int d^{n}x\,\left[ -\frac{1}{4}F_{\mu \nu }^{2}+{\frac{1}{2}}B_{\mu
}B^{\mu }+2K_{\mu }\,A^{\mu }+ \frac{\lambda }{2}\left( \partial _{\mu }A^{\mu
}\right)
^{2}\right] \right\} \times  \nonumber \\
&&\times \Pi _{r=1}^{N}\det \left[ i\partial \!\!\!/\,-\,m\,-\,{\frac{1}{%
\sqrt{N}}}(e\,\pa\,+\,g\,\pb)+J\!\!\!/^{ r }\,\right] .
\end{eqnarray}

Now we compute the  fermionic determinant keeping terms of order $\, (1/N)^0 \, $ and
$(1/N)^{1/2}$. Higher order terms will be neglected. Furthermore, since we are only
interested in two point correlators, the terms higher than quadratic in the sources
$J_{\mu}^r $ will not be taken into account either. This amounts  to the quadratic
approximation for the fermionic determinant which gives
\begin{eqnarray}
Z \left[ J^r_{\mu },K_{\mu }\right] &=&\int \mathcal{D}A_{\mu }\mathcal{D}B_{\mu }
\exp \frac{\imath}{2}\int \frac{d^{D}k}{(2\pi )^{D}}\left\{ -%
{\tilde{A}}_{\mu }\left[ \theta ^{\mu \nu }k^{2}\left( 1-\lambda \right)
+\lambda \,k^{2}\,g^{\mu \nu }\right] {\tilde{A}}_{\nu }\,+\,{\tilde{B}}%
^{\mu }{\tilde{B}}_{\mu }\right.  \nonumber \\
&&\left. \,+\,\sum_{r=1}^{N}(e\,\frac{\tilde{A}_{\mu }\,}{\sqrt{N}}+\,g\,%
\frac{\tilde{B}_{\mu }}{\sqrt{N}}-\tilde{J}^r_{\mu})\Pi ^{\mu \nu
}(k^2)(e\,\frac{\tilde{A}_{\nu }}{\sqrt{N}}\,+\,g\,\frac{\tilde{B}_{\nu
}}{\sqrt{N}}-\tilde{J}^r_{\nu})+\,2\,{\tilde K}_{\mu }\,{\tilde A}^{\mu }\right\}
\label{zb}\;\;,
\end{eqnarray}

\noindent where $\theta ^{\mu \nu }=g^{\mu \nu }-\frac{k^{\mu }k^{\nu }}{%
k^{2}}$ and the tilde over the fields
represent their Fourier transformations in momentum space.
The quantity $\Pi ^{\mu \nu }$ is the vacuum polarization tensor:

\begin{equation}
\Pi ^{\mu \nu }(k)=i\int \;\frac{d^{D}p}{(2\pi )^{D}}\;tr\left[ \frac{1}{\ps %
-m+i\epsilon }\gamma ^{\mu }\frac{1}{(p\!\!\!/+k\!\!\!/)-m+i\epsilon }\gamma ^{\nu
}\right] \label{pimini}.
\end{equation}

In order to proceed further we have to calculate $\Pi ^{\mu \nu }$ which
depends on the dimensionality of the space-time.

%
%

\section{Bosonization from two point  correlators in $D=2$}

In this section we restrict ourselves to the $\, D=2 \, $ case. Using dimensional
regularization we obtain,
below the pair creation threshold ($%
z\equiv \frac{k^{2}}{4\,m^{2}}<1$), :

\be
\Pi_{\mu\nu} \, = \, {\tilde \Pi } (k^2) \, \theta_{\mu\nu} \ee

\no with

\begin{equation}
{\tilde{\Pi}}(k^{2})={\frac{1}{\pi }}\left[ 1-\frac{1}{\left[ z\left(
1-z\right) \right] ^{\frac{1}{2}}}\arctan \sqrt{\frac{z}{1-z}}\right]
;\,\,\,\,\,0<\,z\,<\,1,
\end{equation}

\begin{equation}
{\tilde{\Pi}}(k^{2})={\frac{1}{\pi }}\left[ 1-\frac{1}{2}\,\frac{1}{\sqrt{%
z\,(z-1)}}\ln \left( {\frac{\sqrt{(1-{z})}+\sqrt{-z}}{\sqrt{(1-{z})}-\sqrt{-z%
}}}\right) \right] ;\,\,\,\,\,\,\,z<\,0.
\end{equation}

\no Once the tensor $\Pi %
^{\mu \nu }$ is calculated one is left with a gaussian integral over the vector fields
$A_{\mu }$ and $B_{\mu }$ from which a generating functional  quadratic in the sources
is derived. Such generating functional furnishes the following two point
correlators\footnote{More precisely, we should have written explicitly the two point
functions in the form $<G(k)H(p)>= I(k)\delta^{(2)} (k+p)\, $  but in this article we
will not display the delta function for a matter of convenience. Notice also that when
we write $A_{rs}=F + G \delta_{rs}$ it is assumed that $F $ multiplies a $N\times N$
matrix where all entries are equal to one.}

\begin{eqnarray}
<j_{\mu }^{r }\left( k\right) \,j_{\nu }^{s}\left( -k\right)
> &=& \,-\frac{\tilde{\Pi}^{2}\left( e^{2}- k^{2}g^{2}\right) }{N D}
\,\theta ^{\mu \nu }\,+\,\tilde{\Pi}%
\,\theta ^{\mu \nu }\,\delta ^{rs}\, \label{jj} , \\
<j_{\mu }^{ r }\left( k\right) \,A_{\nu }\left( -k\right) > &=& \,%
\frac{\,e\,\tilde{\Pi}}{\sqrt{N} D }\,\,\,\theta ^{\mu \nu }, \label{jA} \\
<A^{\mu }\left( k\right) \,A^{\nu }\left( -k\right) > &=& \,\left( -\frac{1 }{\lambda
k^{2}}+\frac{1+g^{2}\tilde{\Pi}}{D }\right) \theta ^{\mu \nu }+\frac{g^{\mu \nu }}{%
\lambda k^{2}}\label{AA2},
\end{eqnarray}
\noindent where \be D \, = \, \left[ \tilde{\Pi}\,e^{2}-k^{2}\left(
1+g^{2}\tilde{\Pi}\right) \right]. \ee \no The tensor structure of the above
correlation functions are in full agreement with the corresponding Ward identities
based on the $U\left(1 \right)$ symmetry and it will play a key role in our
 bosonization procedure.
 At this point it is important to stress that our approach deviates
 from that of reference \cite{bm2}.
 The $\, 1/N \, $ expansion we have relied upon, which coincides
 with the quadratic approximation for the fermionic determinant, is not equivalent
 to the second order weak coupling expansion  used in \cite{bm2}.


Now, in order to derive a bosonized expression for the currents
$j_{\mu }^r =\overline{\psi }_r\,\gamma_{\mu }\psi_r $, we write down the most general decomposition for a vector in the momentum space:

\begin{equation}
j_{\beta }^r(k) \, = \, \,\epsilon_{\beta \delta }k^{\delta }\,\phi^r(k) \,+\,k_{\beta
}\,\varphi^r (k).
\end{equation}

\noindent Substituting it in (\ref{jj}) and using the identity $\epsilon _{\beta
\delta }k^{\delta }\epsilon_{\alpha \gamma }k^{\gamma }= k^{2}\theta _{\alpha \beta }$
we obtain
\begin{eqnarray}
<j^r_{\alpha }\left( k\right) \,j^s_{\beta }\left( -k\right) >&=& -k^{2}\theta
_{\alpha \beta }\,<\phi^r\left( k\right) \phi^s \left( -k\right) >-k_{\alpha }k_{\beta
}<\varphi^r \left( k\right) \varphi^s \left( -k\right) > \nn \\
&-&\epsilon _{\alpha \delta }k^{\delta }k_{\beta }<\phi^r\left( k\right) \varphi^s
\left( -k\right) >-\epsilon _{\beta \gamma }k^{\gamma }k_{\alpha }<\varphi^r \left(
k\right) \phi^s\left( -k\right) > \nn \\
&=& -\frac{\tilde{\Pi}^{2}\left( e^{2}- k^{2}g^{2}\right) }{N D}
\,\theta_{\alpha \beta }\,+\,\tilde{\Pi}%
\,\theta_{\alpha \beta}\,\delta ^{rs}.
\end{eqnarray}

\noindent From the above it is not difficult to derive
\begin{eqnarray}
<\varphi^r \left( k\right) \varphi^s \left( -k\right) >&=& 0 , \label{vfvf}\\
<\phi^r \left( k\right) \varphi^s \left( -k\right) > &=& 0 , \label{fvf} \\
<\phi^r \left( k\right) \phi^s \left( -k\right) > &=& -\frac 1{k^2}\left\lbrack
\,-\frac{\tilde{\Pi}^{2}\left( e^{2}- k^{2}g^{2}\right) }{N D}
\,+\,\tilde{\Pi}%
\,\,\delta ^{rs}\right\rbrack , \label{ff} \end{eqnarray}

\noindent from which one can safely set $\varphi =0$. On the other hand, substituting
a general decomposition
\begin{equation}
<\phi \left( k\right) A_{\mu }\left( -k\right) >\,=\,M\,\in _{\mu \delta }k^{\delta
}\,+\,Q\,k_{\mu },
\end{equation}
in the mixed correlation functions $ <j^r_{\mu }( k) A_{\nu }( -k )
> $ given in (\ref{jA}) we conclude that
\begin{equation}
<\phi^r \left( k\right) A_{\mu }\left( -k\right) >\, =\, -\frac{e
\tilde{\Pi}}{\sqrt{N}k^2 D}\,\epsilon_{\mu \delta }k^{\delta };\, \label{fA}
\end{equation}


Now we are in a position to derive the bosonic Lagrangian density $\, {\cal L}_B
(A_{\mu},\phi^r) \, $ which is compatible with the correlation functions
(\ref{AA2}),(\ref{ff}) and (\ref{fA}). For this purpose we start from the following
Ansatz
\begin{equation}
{\cal L}_{B}(A_{\mu},\phi^r) \,=\,\phi^r \,R_{rs}\,\phi^s +2\,S_r\,\phi^r \,\epsilon
_{\mu \nu }k^{\nu }A^{\mu }+A^{\mu }A^{\nu }\left( T_{1}\,\theta _{\mu \nu
}\,+\,T_{2}\,g_{\mu \nu }\right),
\end{equation}
\no where $R_{rs}\, , \, S_r \, , \, T_1 $ and $\, T_2$ will be determined as follows.
We introduce the external sources $X^r $ and $  K_{\mu} $ and define the generating
functional

\be Z_B\left[ X_r,K_{\mu }\right] \, = \, \int \prod_{r=1}^N{\cal D}\phi^r {\cal
D}A_{\mu } e^{i \int d^2x\left[{\cal L}_B(A_{\mu},\phi^r)+ X_r \phi^r + K_{\mu
}A^{\mu}\right]} \label{star}. \ee

\no Assuming that  $R_{rs}$ is a symmetric non-singular matrix we have performed the
gaussian integrals in (\ref{star}) and obtained an explicit formula for $Z_B\left[
X^r,K_{\mu }\right]$ from which the two-point correlators can be obtained. By matching
these correlators with (\ref{AA2}),(\ref{ff}) and (\ref{fA}) we determine the bosonic
Lagrangian uniquely :

\bea {\cal L}_B(A_{\mu},\phi^r)  \, &=& \, \frac 12\left[ \sum_{r=1}^N \phi^r \frac
{k^2}{{\tilde \Pi}}\phi^r + \frac{g^2}{N} \, k^2 \left(\sum_{r=1}^N \phi^r
\right)\left(\sum_{s=1}^N \phi^s\right)\right]  \nn \\
& &  - \frac{A^{\alpha}A^{\beta}}2\left[(1-\lambda )k^2\theta_{\alpha\beta } + \lambda
k^2 g_{\alpha\beta} \right] \nn \\
& & + \frac e{\sqrt{N}} \, \epsilon_{\mu\nu} A^{\mu}k^{\nu} \left( \sum_{r=1}^N \phi^r
\right)\label{A}\quad. \eea

\no All the steps which lead us to (\ref{A}) are technically simple and not very
elucidating. We should only mention that no expansions on either $1/N$ or any of the
coupling constants have been made in those intermediate calculations. Furthermore,
notice that if we quantize ${\cal L}_B(A_{\mu},\phi^r) $ and integrate over the scalar
fields $\phi^r $ in (\ref{A}) this will lead to a nonlocal effective action for the
photon which was studied in \cite{ddh2} where we concluded that, although nonlocal,
the theory is free of tachyons. Next, by comparing with the Lagrangian density written
in terms of fermionic fields in (1)  we have the bosonization formulae for each
fermion flavor (no sum over repeated roman indices below) :

\bea \overline{\psi }_r\,\gamma_{\mu }\psi_r (k)\, &=& \, \epsilon_{\mu\nu} k^{\nu}
\phi^r (k), \label{A1} \\
-\bar{\psi}_{r}(k)\,(k \!\!\!/\,+\,m\, )\psi_{r} (k)\, &=& \, \frac 12 \phi^r \frac
{k^2}{{\tilde \Pi}}\phi^r . \label{A2} \eea

\no Now some comments are in order. First of all, if for some given flavor we do a
$U(1)$ transformation ($\psi_r\to e^{i\alpha } \psi_r $ )  in the expectation value
$<j_{\mu}^r >$ and use any regularization scheme preserving the $U(1)$ symmetry it
will be easy to derive the Ward identity $<j^r_{\mu}\partial^{\nu}j^r_{\nu}> =0 $
which implies the tensor structure $<j^r_{\mu}j^r_{\nu}>\, \propto \,\theta_{\mu\nu} $
and consequently we will have (\ref{A1}). So the current is topological due to the
$U(1)$ global symmetry and that must hold non-perturbatively. On the other hand, the
bosonization rule (\ref{A2}) is only approximate since the full expression would
require, in our approach, the complete knowledge of the fermionic determinant which is
only possible for $m\to 0$. In this case $\tilde{\Pi }\to 1/\pi $ and we end up with
$N$ massless scalar fields topologically coupled to the gauge field. Integration over
the gauge field leads to $N-1$ massless scalar modes and one mode with $\, m^2 =
e^2/(N\pi + g^2) \, $. Thus, reproducing the particular case of the so called
Schwinger-Thirring model for $N=1$ \cite{st}, as well as the Schwinger model result
$m^2=e^2/\pi $ for $g\to 0 $ and $N=1$. In the opposite limit of large mass ($z\to 0$)
we have ${\tilde \pi}\to -2z/3 $ at leading order. Substituting back in (\ref{A}) we
arrive at a divergent result at $m\to \infty $ which is in agreement with the
$m\to\infty$ limit of the corresponding Sine-Gordon model. See \cite{kondo} for a
similar comparison in the case of the massive Thirring model without eletromagnetic
coupling.

\section{Bosonization from two-point correlators in $D=3$}

In $\, D=3 \, $ dimensions the vacuum polarization tensor (\ref{pimini}) calculated by
means of dimensional regularization is given by :

\begin{equation}
\Pi ^{\mu \nu }(k)=i\,\,\,\Pi _{1}\,E^{\mu \nu }+\Pi _{2}k^{2}\,\theta ^{\mu \nu
}\label{pimini3} \, ,
\end{equation}

\noindent with $E^{\mu \nu }\equiv \epsilon ^{\mu \nu \rho }k_{\rho }$ and, in the
range $0\leq \,z\,<\,1$,

\begin{equation}
\Pi _{1} \,=\,-\frac{1}{8\pi z^{1/2}}\ln \left( \frac{%
1+z^{1/2}}{1-z^{1/2}}\right) ;\,\qquad \Pi _{2}=\,\frac{1}{16\pi m\, z}\left[ 1-\left(
\frac{1+z}{2 z^{1/2}}\right)\ln \left( \frac{%
1+z^{1/2}}{1-z^{1/2}}\right)  \right] . \label{p1p2a}
\end{equation}
While for  $\, z<0 \, $ we have

\begin{equation}
\Pi _{1} \,=\, -\frac{1}{4\pi (-z)^{1/2}}\arctan \sqrt{-z};\, \,\, \,\, \Pi
_{2}=\,\frac{1}{16\pi m\, z}\left[ 1-\left( \frac{1+z}{2 (-z)^{1/2}}\right)\arctan
\sqrt{-z}\right] . \label{p1p2b}
\end{equation}

\no In fact the $\Pi_{1}$ amounts to a regularization dependent finite term
\cite{coste,luscher}, which was taken equals to zero due the dimensional
regularization used.

Substituting (\ref{pimini3}) in the general expression (\ref{zb}) we can obtain the
two-point functions:

\begin{eqnarray}
<j_{\mu }^{r }( k) \,j_{\nu }^{s}(-k)
> &=& -\frac 1N \left[k^2\left(\Pi_2 +\frac P{\tilde Q} \right)
\,\theta_{\mu \nu }\,+\, i\,\Pi_1\left(1-\frac{k^2}{\tilde Q}\right) E_{\mu\nu} \right] \nn \\
&+& \left( k^2\Pi_2 \theta_{\mu\nu} + i\,\Pi_1 E_{\mu\nu} \right) \delta_{rs} , \label{jj3}\\
<j_{\mu }^{r}\left( k\right) \,A_{\nu }\left( -k\right) > &=& \,%
\frac{e}{\tilde{Q}\sqrt{N} }\,
\left[P\theta _{\mu \nu }- i\,\Pi_1 E_{\mu\nu} \right] , \label{jA3} \\
<A_{\mu }\left( k\right) \,A_{\nu }\left( -k\right) > &=& \frac{g_{\mu\nu}}{\lambda
k^{2}}+ \left[\frac{e^2P}{k^2\tilde{Q}}-\frac{(1+\lambda )}{\lambda k^2}
\right]\theta_{\nu\mu} -\frac{i e^2\Pi_1}{k^2\tilde{Q}}E_{\mu\nu} , \label{AA3}
\end{eqnarray}

\no where we found convenient to define

\bea
P\, &=& \, (e^2-k^2 g^2)(k^2\Pi_2^2-\Pi_1^2)-k^2 \Pi_2 \quad , \label{P} \\
\tilde{Q} \, &=& \, k^2\left[(e^2-k^2 g^2)\Pi_2 -1\right]^2 - (e^2-k^2 g^2)^2\Pi_1^2
\quad . \label{tildeQ} \eea

\noindent In Analogy to the $D=2$ case, we use for the $U(1)$ current a general
decomposition in the momentum space :

\begin{equation}
j^r_{\alpha }(k) \, = \,\epsilon _{\alpha \beta \gamma }k^{\beta }\,B_r^{\gamma
}(k)\,+\,k_{\alpha }\,\phi^r (k), \label{d3} \end{equation}

\noindent from which we get

\bea <j^r_{\alpha }\left( k\right) \,j^s_{\beta }\left( -k\right) >&=&-k_{\alpha
}k_{\beta }<\phi^r \left( k\right) \phi^s \left( -k\right)
>+\,E_{\alpha \gamma }E_{\delta \beta
}<B_r^{\gamma }\left( k\right) B_s^{\delta }\left( -k\right) > \nn \\
&+&\,k_{\alpha }\,E_{\beta \delta }\,<\phi^r \left( k\right) B_s^{\delta }\left(
-k\right) >+k_{\beta }\,E_{\alpha \gamma }<B_r^{\gamma }\left( k\right) \phi_s \left(
-k\right) >. \,\eea

\no Multiplying the last expression by $k^{\alpha}k^{\beta}$ we conclude that
$<\phi^r\left( k\right)\phi^s \left(- k\right)> = 0 $. Now, multiplying the resulting
expression by $k^{\alpha }$ we have $E_{\alpha\beta }<\phi^r(k) B_s^{\beta}(-k)>=0$. A
similar manipulation was used in the last section to derive (\ref{vfvf}),(\ref{fvf})
and (\ref{ff}). Concluding, we can certainly neglect the scalar fields $\phi^r = 0 $
and minimally bosonize the $U(1)$ current in $D=3$ in terms of a bosonic vector field.
The bosonic version of the current is once again of topological nature and identically
conserved. As in $D=2$ case this happens because of the $U(1)$ global symmetry of the
fermionic Lagrangian. Taking $\phi^r=0$ and substituting the decomposition (\ref{d3})
in (\ref{jj3}) and (\ref{jA3}), after some trivial manipulations, we end up with \bea
<B_r^{\gamma}(k)B_s^{\delta}(-k)> \, &=& \, -C_{rs}g^{\gamma\delta} +
\left(C-\frac{H_2}{k^2}\right)_{rs}\theta^{\gamma\delta}-\frac{(H_1)_{rs}}{k^2}E^{\gamma\delta}
, \label{BB}\\
<A^{\nu}(k)B_s^{\mu}(-k)> \, &=& \,
-\left(\frac{ie\Pi_1}{\sqrt{N}\tilde{Q}}+D\right)_s g^{\nu\mu} - \left(\frac{e
P}{\sqrt{N}\tilde{Q}k^2}\right)_s E^{\nu\mu}+ D_s \theta^{\nu\mu} , \label{AB} \eea

\no where $C_{rs}$ and $D_s $ are arbitrary and

\bea (H_1)_{rs} \, &=& \, i\,\Pi_1 \,\delta_{rs} + \frac{i\,\Pi_1}N
\left(\frac{k^2}{\tilde{Q}}-1
\right) , \label{H1} \\
(H_2)_{rs} \, &=& \, k^2\Pi_2 \,\delta_{rs} - \frac{k^2}N\left( \Pi_2 +\frac
P{\tilde{Q}} \right). \label{H2} \eea

\no The arbitrariness of $ C_{rs} $ and $D_s$ shows that the bosonization fields
$B_r^{\mu}$ must be gauge fields but the corresponding gauge symmetry is independent
of the eletromagnetic one. Analogous to the last section, we next derive a Lagrangian
density compatible with the two-point correlators (\ref{BB}),(\ref{AB}) and
(\ref{AA3}). Let us suppose we have a bosonic Lagrangian density ${\cal
L}_B(A_{\mu},B^{\nu}_r) $ of the form

\be {\cal L}_{B}(A_{\mu},B^{\nu}_r)\, = \,B_r^{\mu }\, {\cal O}_{\mu\nu}^{rs}
\,B_s^{\nu } + u^s B_s^{\mu }\,E_{\mu \nu }\,\,A^{\nu }+A_{\mu }A_{\nu }\left(v_1 \,
\theta_{\mu\nu} + v_2 g_{\mu\nu} \right) , \label{lb3} \ee

\no Where $\,  {\cal O}_{\mu\nu}^{rs} \, $ has a  general tensor structure in both
momentum and flavor space,

\be {\cal O}_{\mu\nu}^{rs} \, = \, a^{rs}\, \theta_{\mu \nu }+ b^{rs}\, E_{\mu \nu } +
d^{rs}\, g_{\mu \nu } \label{operator}\ee

 \noindent The quantities $a^{rs},b^{rs},d^{rs},u^s, v_1, v_2 $ will
be determined by matching the two-point correlators  derived from the generating
functional :

\be Z_B\left[ Y_{\nu}^r,K_{\mu }\right] \, = \, \int  \prod_{r=1}^N{\cal
D}B^{\mu}_r{\cal D}A_{\mu } e^{i \int d^2x\left({\cal L}_{B}(A_{\mu},B^{\nu}_r) +
Y^r_{\mu} B_r^{\mu } + K_{\mu} A^{\mu}\right)}, \label{star3}\ee

\no with the correlators (\ref{BB}),(\ref{AB}) and (\ref{AA3}). Assuming that the
matrices $a^{rs},b^{rs},d^{rs}$ are symmetric and non-singular, our first step  is to
perform the gaussian integrals over the $N$ vector fields  $B^{\nu}_r$ which
furnishes:

\be Z_B\left[ Y_{\nu}^r,K_{\mu }\right]  =  \int {\cal D}A_{\mu } e^{i \int
d^2x\left[-\frac 14 \left( Y_r^{\mu}  + u_r A_{\beta}E^{\beta\mu}\right)({\cal
O}^{-1})^{rs}_{\mu\nu}  \left( Y_s^{\nu}  + u_s E^{\nu\beta}A_{\beta}\right) + A^{\mu
}A^{\nu }\left(v_1 \, \theta_{\mu\nu} + v_2 g_{\mu\nu} \right) + K_{\mu}
A^{\mu}\right]}. \label{star4}\ee

\no The inverse operator ${\cal O}^{-1}$ has the same structure of (\ref{operator}): $
({\cal O}^{-1})_{\mu\nu}^{rs} =  {\tilde a}^{rs} \theta_{\mu \nu}+ {\tilde b}^{rs}
E_{\mu \nu } + {\tilde d}^{rs}\, g_{\mu \nu } $ where the tilde variables are
functions of the non-tilde ones. In particular, ${\tilde d}_{rs}=(d^{-1})_{rs} $.
 Since we have the contractions $E^{\mu\nu}E_{\nu\alpha}=-k^2\,
\theta^{\mu}_{\alpha}\, $; $\, E^{\mu\nu}\theta_{\nu\alpha}= E^{\mu}_{\alpha}$ it
becomes clear from (\ref{star4}), with a little thought, that even after integrating
over $A_{\mu}$ the only term in $log\, Z_{B}$ quadratic in the sources $Y^{\mu}_{r}$
and contracted via the metric tensor is : $\, - \frac 14\, Y^{\mu}_{r} Y^{\nu}_{s}
{\tilde d}^{rs} g_{\mu \nu}\, $. Therefore the two point functions (\ref{BB}) will
require the identification $C^{rs}=({\tilde d}/2)^{rs}= (d^{-1}/2)^{rs}$.

 Given that
the  matrix $C^{rs}$ is arbitrary, in order to simplify the calculations, we have
chosen it proportional to the identity  and consequently we reduce the  undetermined
parameters  in our original Ansatz for the bosonic lagrangian  since $d^{rs}=d\,
\delta^{rs}$. For analogous reasons, after integrating over $A_{\mu}$, the crossed
terms in the sources $Y^{\mu}_{r}$ and $K^{\nu}$ will be only contracted by tensors
$E_{\mu \nu}$ and $\theta_{\mu \nu}$. Thus, the term proportional to $g^{\nu \mu}$ on
the right handed side of the correlation functions (\ref{AB}) must vanish, which fixes
$D=-\frac{ie\Pi_1}{\sqrt{N}\tilde{Q}}$. Finally, comparing the other parts of the
correlators  (\ref{BB}), (\ref{AB}) and (\ref{AA3}) with the final expression for $\,
\log Z_B\, $ one can determine the  quantities $a^{rs}\, , \, b^{rs}$ and $u^s$ as
functions of ``d''  which remains unfixed. We spare the reader the lengthy details
since they are totally technical and not very illuminating. We only stress that no
further simplification hypothesis is used and no expansion is made in neither $1/N$
nor in any of the coupling constants. In those intermediate steps all correlators are
treated as if they were exact.

In order to make the final answer for the bosonic Lagrangian density  more familiar we
have redefined the unfixed parameter $d\equiv -{\tilde \lambda}k^2/2 $. In the
momentum space we have:

\bea {\cal L}_B(A_{\mu},B_r^{\alpha})  &=&
-\frac 12\left[ \sum_{r=1}^N  \frac {B^{\alpha}_r\left(k^2 \Pi_2\,
\theta_{\alpha\beta} + i\, \Pi_1 \epsilon_{\alpha\beta\gamma}k^{\gamma}\right)B^{\beta}_r }
{k^2 \Pi_2^2 - \Pi_1^2}+ \frac{g^2}{N} \, \left(\sum_{r=1}^N
B^{\alpha}_r
\right)k^2\, \theta_{\alpha\beta}\left(\sum_{s=1}^N B^{\beta}_s\right)\right]  \nn \\
& &  - \frac{A^{\alpha}A^{\beta}}2\left[(1-\lambda )k^2\theta_{\alpha\beta } + \lambda
k^2 g_{\alpha\beta} \right]+ \frac e{\sqrt{N}} \,
 \epsilon_{\mu\nu\gamma} k^{\gamma}\, A^{\mu}  \sum_{r=1}^N B^{\nu}_r
\nn \\
& & + \sum_{r=1}^N\frac{B_r^{\alpha}B^{\beta}_r}2\left[\tilde{\lambda}
k^2\theta_{\alpha\beta } - \tilde{\lambda} k^2 g_{\alpha\beta} \right] \label{A3}\quad
. \eea

\no Notice that $\lambda $ and $\tilde{\lambda}$ are interpreted as independent gauge
fixing parameters such that the generating functional  (\ref{star3}) reproduces the
correlation functions (\ref{AA3}),(\ref{BB}) and (\ref{AB}) with

\bea C_{rs}\, &=& \, -\frac 1{{\tilde\lambda}k^2}\, \delta _{rs} \label{C}\\
D \, &=& \, -\frac{ie\Pi_1}{\sqrt{N}\tilde{Q}} \label{D} \eea

\no Integration over the vector fields $B^{\alpha}_r$ will lead to an effective
eletromagnetic theory which was also studied in \cite{ddh2} ( see also \cite{ddh1} )
where we have analysed its pole structure and concluded that no tachyons appear just
like in its two dimensional couterpart (\ref{A}).

Comparing (\ref{A3}) with the original Lagrangian
(1) written in terms of fermionic fields
we have the bosonization rules for each fermion flavor ($\, r=1,2,\cdots , N \, $):

\bea \overline{\psi }_r\,\gamma_{\mu }\psi_r (k)\, &=& \, \epsilon_{\mu\nu\alpha} k^{\nu}
B^{\alpha}_r(k) , \label{A13} \\
-\bar{\psi}_{r}(k)\,(k \!\!\!/\,+\,m\, )\psi_{r} (k)\, &=& \, -\frac 12 B^{\alpha}_r
\frac {\left(k^2 \Pi_2\, \theta_{\alpha\beta} + i\, \Pi_1 E_{\alpha\beta}\right)} {k^2
\Pi_2^2 - \Pi_1^2}B^{\beta}_r . \label{A23} \eea

Similarly to the $D=2$ case, we have obtained (\ref{A13}) from the fact that for each
fermion flavor we have the tensor structure $<j_{\mu}j_{\nu}> \, \propto a\,
\theta_{\mu\nu} + b \, E_{\mu\nu} $, which has been derived by calculating the
fermionic determinant up to the quadratic order. However, this tensor structure must
hold beyond that approximation ( non-perturbatively ) as a consequence of the Ward
identity $<j_{\mu}\partial^{\nu}j_{\nu}>=0\, $ which must be true also in $D=3$ by
using a gauge invariant regularization. On the other hand, the bosonization rule
(\ref{A23}) holds as a consequence of the quadratic approximation for the fermionic
determinant. Taking $\, m\to 0 \, $ (or $\, z\to -\infty\, $) in the expressions
(\ref{p1p2b}) we have $\Pi_1\to 0 $ and $\Pi_2\to 1/(16\sqrt{-k^2})$. Plugging back in
(\ref{A13}) and (\ref{A23}) we recover the result of \cite{marino} with $\beta=1/4$
and $\theta=0$ in that reference which deals with massless fermions. Our rules
(\ref{A13}) and (\ref{A23}) also agree with the case of massive fermions treated in
\cite{bfo}. We stress however, that in both \cite{marino} and \cite{bfo} only free
fermions have been considered while our results have been derived for an interacting
theory with Thirring and eletromagnetic couplings. This indicates that our
bosonization rules are rather universal at the quadratic level. In the opposite limit
of heavy fermions $\, (k^2/4m^2)\, =\, z\to 0 \, $ in (\ref{p1p2a}) we have, at
leading order, $\Pi_2\to 0 $ and $\Pi_1\to -1/(4\pi )$ which leaves us, see
(\ref{A23}), with a bosonic local and topological Chern-Simons Lagragian for the field
$\, B_r^{\mu} \, $ in agreement with the findings of \cite{schaposnik} (see also
\cite{banerjeea} and \cite{banerjee1} for higher orders ).

\section{Conclusions}

We have derived bosonization maps
 for the $U(1)$ currents and the fermion Lagrangian densities for both
$QED_{2}$ and $QED_{3}$ with $N$ fermion flavors and Thirring self-interaction. Both
results hold for finite fermion masses and no derivative expansion on $\frac km $ is
made as in \cite{sghosh}. By turning off the interactions we can reproduce the results
of \cite{bfo} for a particular choice of the regularization parameter in that
reference. Our calculations show that the $U(1)$ currents, when written in terms of
bosons, must be identically conserved (topological) as a direct consequence of the
Ward-identity $\, <j_{\mu}\partial^{\nu}j_{\nu}>\, =\, 0  $ which is true
non-perturbatively if there is no $U(1)$ anomaly. In particular, there is no need for
looking at higher point functions  or compute the fermionic determinant beyond the
quadratic approximation to confirm that. On the other hand, our maps for the
Lagrangian densities (\ref{A2}) and (\ref{A23}), which are also independent of the
interactions, hold  only perturbatively  due to our quadratic approximation for the
fermionic determinant which amounts to neglect terms of order higher than $(1/N)^{1/2}
\,$ in the fermionic determinant and consider at most two point current correlators to
derive the bosonization rules.

Finally, we should mention that the results derived here could have been obtained by a
technically more direct way along the lines of \cite{bos} (see also \cite{bq}) but the
approach used here does not use any auxiliary field and clarifies the fundamental role
of the two point correlators. Besides, it might be also useful (work in progress) for
deriving approximate bosonic maps for other fermion bilinears like the mass term $\,
{\overline\psi}\psi\, $ in $D=3$, which apparently could not be done by using the
approach of \cite{bos}.

\section{Acknowledgements}

This work was partially supported by \textbf{CNPq} and \textbf{FAPESP},
Brazilian research agencies.

\newpage

\end{document}